# The Effect of Land Albedo on the Climate of Land-Dominated Planets in the TRAPPIST-1 System

Andrew J. Rushby,[1] Aomawa L. Shields,[1] Eric T. Wolf,[2] Marysa Laguë,[3] and Adam Burgasser[4]

[1]*Department of Physics & Astronomy, University of California, Irvine*
*4129 Frederick Reines Hall, Irvine, CA. 92697-4575*
[2]*Laboratory for Atmospheric and Space Physics, University of Colorado, Boulder*
*3665 Discovery Drive, Campus Box 600, Boulder, CO. 80303*
[3]*Department of Earth and Planetary Science, University of California, Berkeley*
*307 McCone Hall, Berkeley, CA. 94720-4767*
[4]*Department of Physics, University of California, San Diego*
*9500 Gilman Drive, La Jolla, CA. 92093*

## ABSTRACT

Variations in the reflective properties of the bulk material that comprises the surface of land-dominated planets will affect the planetary energy balance by interacting differently with incident radiation from the host star. Furthermore, low-mass cool stars, such as nearby M8V dwarf TRAPPIST-1, emit a significant fraction of their flux in longer wavelengths relative to the Sun in regions where terrestrial materials may exhibit additional variability in albedo. Using the Community Earth System Model (CESM) we investigate the effect of the composition of the land surface and its albedo on planetary climate in the context of spatially homogeneous, entirely land-covered planets with dry atmospheres at the orbital separation of TRAPPIST-1d, TRAPPIST-1e, and TRAPPIST-1f. We use empirically derived spectra of four terrestrial compositional endmembers (granite, calcite, aridisol, and dune sand) and a composite spectrum of TRAPPIST-1 for these simulations and compare these model output to an aquaplanet and several Sol-spectrum control cases. We report a difference of approximately 50 K in global mean surface temperature, variations in atmospheric rotational features, and a reduction in cross-equatorial heat transport between scenarios in which materials with higher albedo in the infrared (calcite and dune sand) were used and those with more absorptive crustal material, such as granite or dry soils. An aquaplanet TRAPPIST-1d scenario results in an unstable runaway greenhouse regime. Therefore, we demonstrate that determining the composition and albedo of continental landmasses is crucial for making accurate determinations of the climate of terrestrial exoplanets.

## 1. INTRODUCTION

Recent work in the broad field of planetary climatology has identified a number of feedback processes within the planet system that provide a non-linear response to perturbations of the climate. These feedbacks include the carbonate-silicate feedback, which acts as a buffering control on the surface temperature and atmospheric $CO_2$ inventory through the action of continental weathering and volcanic outgassing (Walker et al. 1981; Edson et al. 2012; Rushby et al. 2018). The phenomenon described as the ice-albedo feedback, which affects the energy balance of the planet through the interaction between instellation (incident starlight), ice distribution and salinity, and surface albedo, has also received recent modelling attention (Shields et al. 2013; Shields & Carns, 2018; Rushby et al. 2019).

Corresponding author: Andrew J. Rushby
arushby@uci.edu

These feedbacks have been conceptually proven to be powerful controls on planetary climate and emphasize the eventual need for high spectral, temporal, and spatial resolution data from terrestrial planets when making assessments of their potential habitability (Cockell et al. 2016). At this stage, many aspects of the terrestrial planet system, and especially so in the case of distant exoplanets, can only be investigated using computer modelling (Shields, 2019).

There has been relatively little investigation of the effect of land albedo in modulating non-linear responses to external forcings of the climate system. Land albedo describes the reflectance of some fraction of incident radiation by the material comprising the land surface of the planet, and varies according to the composition of the material as well as the spectral energy distribution (SED) of the host star. Previous work has demonstrated the importance of this spectral dependence of albedo, as well as the amount and spatial distribution of the continental crust, on surface temperature and climate stability (Abe et al. 2011; Leconte et al. 2013; Shields et al. 2013; Foley 2015). Rushby et al. (2019) found evidence of a potential reversal of the ice-albedo feedback on land-dominated planets (land fraction ($F_L$) > 0.7) orbiting M-dwarf stars due to the interaction between the host star's SED, which peaks in the near-infrared, and the optical/thermal properties of water ice, which is more absorptive than land in these wavelengths. However, these studies focused on the effect of the changing SED of the host star, and the variable land surface fraction (i.e. how much land versus ocean is exposed to instellation) on the climate of the planet, often by comparing experimental results to that of a completely ocean-covered ('aquaplanet') configuration as a control. The effect of changes in land surface albedo, caused by, for example, variations in the composition of the surface or its overlying cover, in the context of completely or nearly completely land-covered planets ('landplanets') has not yet been explored.

The expected number of terrestrial ($R_p < 1.6 R_\oplus$) planets at habitable zone (HZ)-like orbital separations is expected to grow with more planet discoveries from currently ongoing missions, such as the Transiting Exoplanet Survey Satellite (*TESS*), as well as those reported by near-future space-based observatories like the James Webb Space Telescope (Ricker et al. 2014; Greene et al. 2016; Kalirai 2018). Next-generation space-based instruments (e.g. the Large UV/Optical/Infrared Surveyor (*LUVOIR*)) and ground-based Extremely Large Telescopes will allow for the characterization of the atmospheres of hundreds of exoplanets through transmission spectroscopy, and for some constraints to be placed on their potential climate and habitability (Kiang et al. 2018; Shields, 2019). However, the relatively recent acceleration of planet discovery also provides observers and theorists alike with the challenge of prioritizing certain candidates or confirmed planets for follow-up observations with time and resource-limited instruments.

The TRAPPIST-1 system, a compact series of seven approximately Earth-sized ($0.772 < R_p/R_\oplus < 1.127$) planets discovered orbiting the nearby, ultra-cool dwarf star TRAPPIST-1 ($0.08 M_\odot$; $0.117 R_\odot$), provides an unprecedented observational and theoretical opportunity to characterize and study the climates of terrestrial planets within a natural laboratory of independent stellar and planetary evolution (Gillon et al. 2017; Grimm et al. 2018). Observed masses and densities of these planets suggest that TRAPPIST-1e (as well as planet c) are likely 'rocky', whereas the mass-radius scaling for TRAPPIST-1d (as well as planets b, f, g, and h) suggests that it likely has a thick, volatile-rich atmosphere or is dominated by oceans or ice (Grimm et al. 2018). However, due to uncertainties in the densities and water content of these planets some uncertainties in their compositions remain (Unterborn et al. 2018). Since the discovery of this system, extensive modelling efforts using a range of climate models, including 1-D energy balance models (EBMs) and more computationally intensive 3D global circulation models (GCMs), have been focused on these planets in order to provide tentative estimates of their ability to support an Earth-like biosphere and to determine potential observational biosignatures (Wolf 2017, 2018; Turbet et al. 2018; Kiang et al. 2018).

In this paper, we investigate the effect of land albedo on the climate of TRAPPIST-1d, TRAPPIST-1e, and TRAPPIST-1f (henceforth, planet *d*, *e*, and *f*) by simulating the climates of land-covered planets receiving the amount of instellation received by these planets from a star with the SED of TRAPPIST-1, as well as a control case in which planet *e* is simulated with a solar spectrum (Chance & Kurucz, 2010). Four representative land surface compositions and their high-resolution spectra were used to initialize 3D global circulation model simulations of *d*, *e*, and *f* landplanets. We compare differences in surface temperature, albedo, and planetary heat transport between these four compositional endmembers for each planet and discuss potential implications for their climates and comparative terrestrial planetology.

## 2. GENERAL CIRCULATION MODEL

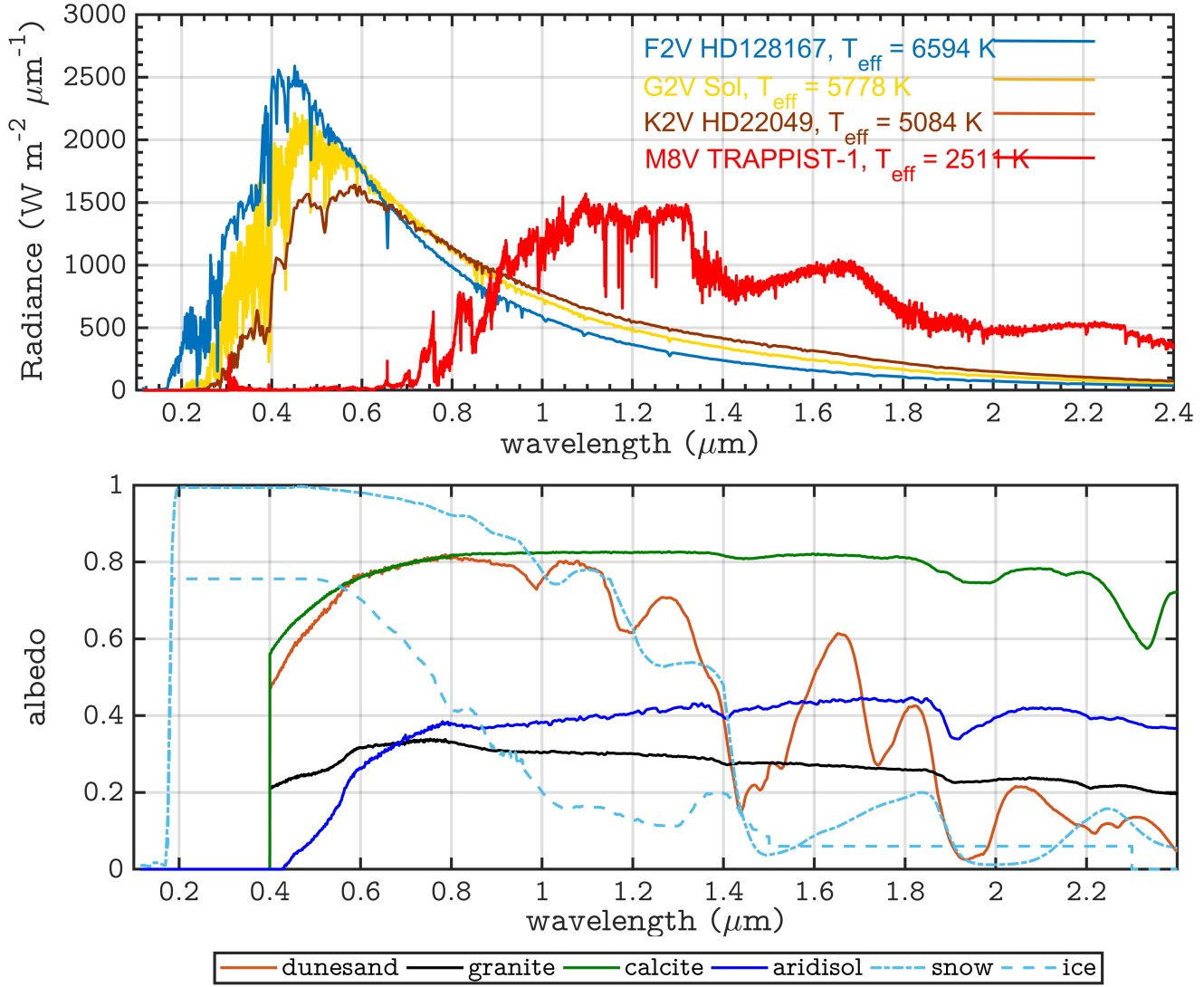

**Figure 1.** Top: Spectral energy distribution (SED) of TRAPPIST-1 (red) (Lincowski et al., 2018), compared to those of standard main sequence F- and K- dwarf stars (Segura et al., 2003) (not used in this work and presented here for comparison only). The solar spectrum, which was used to initialize some control runs, is from Chance & Kurucz (2010). See Rushby et al. (2019) or Shields et al. (2013) for details regarding the reduction of these spectra. Bottom: the spectral distribution of the surface compositions used in this work: dunesand (brown/red), granite (black), calcite (green), aridisol (blue), as well as fine-grained snow (dash-dotted light blue) and blue marine ice (dashed light blue).

We use version 1.2.1 of the National Center for Atmospheric Research (NCAR) Community Earth System Model (CESM), a fully-coupled 3D global circulation model (GCM), in concert with the ExoCAM package[1] which is based on the Community Atmosphere Model (CAM) version 4 and allows for the modification of certain atmospheric parameters for generalized exoplanet applications. ExoCAM employs a flexible, two-stream correlated-$k$ radiative transfer package[2] presently including $H_2O$, $CO_2$, and $H_2$. Our horizontal grid is 4° x 5° and the model atmosphere is comprised of 40 layers with a model top of ∼1 hPa. We used a composite SED derived from observations and models of the nearby M8V ultra-cool dwarf TRAPPIST-1 supplied by the VPL spectral database[3] (originally published in Lincowski et al., 2018). The spectra provides the stellar flux of TRAPPIST-1 from the UV into the MIR at a distance of 1 AU (figure 1, top panel) and is the product of photometric observations and a 2500 K, [Fe/H]=0.0, log g = 5.0 spectral model from the PHOENIX v2.0 spectral database spanning 0.25 to 5.5$\mu$m (Lincowski et al 2018). We also leverage a Solar spectrum in order to act as a control case and to identify any spectral dependence in our results that may have implications for ultra-cool dwarfs and low mass stars (Chance & Kurucz, 2010).

---

[1] https://github.com/storyofthewolf/ExoCAM

We carried out coupled CESM simulations for three terrestrial planets in this system, TRAPPIST-1d ($R_\oplus$ = 0.772±0.002; $M_\oplus$ = 0.41±0.003), TRAPPIST-1e ($R_\oplus$ = 0.918±0.003; $M_\oplus$ = 0.62±0.008) and TRAPPIST-1f ($R_\oplus$ = 1.045±0.003; $M_\oplus$ = 0.68±0.008) (Gillon et al. 2017). We have assumed that all three planets are spin-orbit locked to the host star in circular, short period orbits with zero obliquity (Grimm et al. 2018; Gillon et al. 2017); that is, a single side of the planet is continually facing the star (the substellar/day-side hemisphere), while the other side of the planet is permanently dark (the antistellar/night-side hemisphere). We assume that the planets are rocky and that they have an $N_2$-dominated, 1 bar atmosphere with present Earth atmospheric level (PAL) of $CO_2$ (400 ppmv) and $CH_4$ (1.7 ppmv), overlying a topographically and compositionally uniform land surface (in the case of land-planet simulations). In these cases, the atmospheric and soil water mixing ratios are also set to be very low ($\sim 10^{-12}$), but cannot be zero due to the nature of mass conservation in the Community Land Model (CLM) that represents the land/atmosphere boundary interface for this configuration of CESM. CLM was initialized with the land cover set to be bare ground and with soil radiative properties that vary between experiments.

We also include a configuration in which we simulate the climate of planet $e$ across a range of land surface compositions using a Solar spectrum (see Rushby et al. 2019 for details on the reduction of this photometry) while maintaining other planetary properties to allow us to identify differences in climatic response that may be attributable to spectral differences between host star (denoted as $e_\odot$). Due to computational constraints we limit our investigation into these differences by carrying out only an aquaplanet, granite, and calcite simulation for planet $e$ with a Sol spectrum. Furthermore, we include non-CLM aqua-planet (i.e. entirely ocean-covered) configurations in which we assume **a moist atmosphere and** a well-mixed, 50m deep 'slab' ocean with zero ocean heat transport (see e.g. Shields et al. 2014, Poulsen et al. 2001, Pierrehumbert 2011; Kopparapu et al. 2017). Where sea ice can form, its growth and distribution are controlled by the coupled Los Alamos sea-ice model (CICE) version 4. Ice properties are modified to reflect those appropriate for the TRAPPIST-1 system. We selected four land surface types to represent the spatially uniform model land surface – aridisol, calcite, granite, and dunesand – and obtained their spectra from the NASA/JPL ECOSTRESS spectral library (Meerdink et al 2019; Balridge et al 2009). The dunesand spectrum[4] was derived from a white gypsum sand originating from eolian evaporites; for granite[5] we use a felsic igneous sample consisting of feldspar and quartz with some impurities. The calcite spectrum[6] is derived from a medium grained sample of calcium carbonate, and the aridisol data[7] originates from directional hemispheric reflectance measurements of a xerollic light yellowish brown loam. The albedos of each of these soil types are imposed on the bareground land type in CLM, which is used to uniformly represent all land surfaces. These surface compositions were chosen to provide wide spectral coverage and variability across the visible and near-infrared, while also remaining representative of possible land surface compositions on land-dominated, dry terrestrial planets (see figure 1, bottom panel). Note the relatively wavelength-independent albedo of calcite, granite, and aridisol, compared to the complex reflectance features of the dunesand spectrum. Using the TRAPPIST-1 and surface composition spectra, we were able to generate unique two-band albedos (that represents the total incident radiation scattered from the surface at all wavelengths and all phase angles), required for CESM, for the four compositional endmembers at TRAPPIST-1, as well as for ice and snow 1. The ice albedo describes a representative surface comprised of a 1:1 mixture of course-grained ice and fine-grained snow, overlying either land or ocean surfaces, as used in Rushby et al. (2019) and originally computed by

---

[2] https://github.com/storyofthewolf/ExoRT
[3] http://vpl.astro.washington.edu/spectra/stellar/trappist1.htm
[4] https://speclib.jpl.nasa.gov/ecospeclibdata/soil.entisol.torripsamment.none.all.0015.jhu.becknic.spectrum.txt
[5] https://speclib.jpl.nasa.gov/ecospeclibdata/rock.igneous.felsic.solid.all.granite_h5.jhu.becknic.spectrum.txt
[6] https://speclib.jpl.nasa.gov/ecospeclibdata/mineral.carbonate.none.medium.vswir.c-3a.jpl.beckman.spectrum.txt
[7] https://speclib.jpl.nasa.gov/ecospeclibdata/soil.aridisol.calciorthid.none.all.84p3721.jhu.becknic.spectrum.txt

Shields et al. (2013). For comparison we also used CESM's default ice albedo values (normalized for a planet orbiting a solar-type star) for a second aquaplanet simulation of TRAPPIST-1e to act as a control case. Global mean albedo for the TRAPPIST-1 planets is calculated as the ratio of shortwave upwelling and downwelling radiation, which, by definition, prevents the calculation of albedo on the unlit hemisphere. The top-of-the-atmosphere (TOA) albedos shown in figure 2 have some small zenith angle dependence due to the increased path lengths through the atmosphere at the horizon, balanced between extra shortwave absorption by $CO_2$ and extra scattering. The model shortwave stream includes near-IR component of the radiation from the star, and its reflection, out to ∼12 $\mu$m.

Table 1. Two channel albedo (VIS/IR) values for various surface compositions used in these simulations.

|  | Aridisol | Calcite | Dunesand | Granite | Ice | Snow |
|---|---|---|---|---|---|---|
| TRAPPIST-1 planets | 0.230/0.405 | 0.732/0.802 | 0.718/0.508 | 0.294/0.277 | 0.682/0.134 | 0.978/0.418 |
| Aquaplanet default | - | - | - | - | 0.68/0.91 | 0.3/0.63 |

The modified version of CESM's Community Atmosphere Model (CAM4) module, ExoCAM, divides the incident stellar (shortwave) radiation into 23 wavelength bands between 0.2 and 12.2 $\mu$m. Surface albedo values are split in the model between a shortwave (VIS) and longwave (IR) component at 0.78 $\mu$m, and these are further subdivided into direct and diffuse streams 1. We set the incident flux at the mean orbital separation of TRAPPIST-1d, TRAPPIST-1e, and TRAPPIST-1f to be 1554.48 Wm$^{-2}$ (1.143 $S_\oplus$), 900.32 Wm$^{-2}$ (0.662 $S_\oplus$), and 519.52 Wm$^{-2}$ (0.382 $S_\oplus$) respectively, following the published values in Gillon et al. (2017). Due to the nature of the radiative scheme employed by CAM4, which is optimized for the Sun and cuts the visible and NIR bands neatly at 5$\mu$m, we integrate the near-IR albedo to longer wavelengths in the shortwave radiation calculation in order to capture all the reflected stellar radiation from TRAPPIST-1, but hold the surface emissivity at 1.0 for all wavelengths >5$\mu$m (which corresponds to 0.04% of the total flux from the star). Simulations were performed for 80 to 90 model years, until thermal equilibrium was reached, and we present a 10-year average of the equilibrated data in this work unless otherwise stated.

## 3. RESULTS

### 3.1. *Surface Temperature*

Table 2. Summary of Output from 3D GCM Landplanet Simulations

|  | Global $T_S$ (K) | Mean Dayside $T_S$ (K) | Mean Nightside $T_S$ (K) | TOA Albedo | Energy Flux Potential (PW) |
|---|---|---|---|---|---|
| **TRAPPIST-1d** | | | | | |
| Aridisol | 235.1 | 264.8 | 203.8 | 0.366 | |
| Calcite | 193.5 | 211.9 | 174.1 | 0.714 | -3.3 |
| Dune Sand | 227.3 | 254.6 | 198.5 | 0.456 | |
| Granite | 244.5 | 276.5 | 210.6 | 0.256 | -7.4 |
| **TRAPPIST-1e** | | | | | |
| Aridisol | 207.6 | 232.3 | 181.4 | 0.368 | |
| Calcite | 170.6 | 185.7 | 154.6 | 0.719 | -1.98 |
| Dune Sand | 199.7 | 222.4 | 175.7 | 0.459 | |
| Granite | 215.9 | 242.8 | 187.3 | 0.257 | -4.4 |
| **TRAPPIST-1e$_\odot$** | | | | | |
| Calcite | 172.4 | 188.5 | 155.3 | 0.746 | -2 |
| Granite | 212.1 | 238.1 | 184.5 | 0.367 | -4.2 |
| **TRAPPIST-1f** | | | | | |
| Calcite | 151.7 | 162.7 | 140.2 | 0.717 | -1.26 |
| Granite | 192.5 | 213.5 | 170.2 | 0.258 | -2.88 |

NOTE—TRAPPIST-1e$_\odot$ refers to the simulations in which TRAPPIST-1e was modelled using a solar spectrum. For TRAPPIST-1e$_\odot$ and TRAPPIST-1f only granite and calcite simulations were carried out for computational efficiency (see text). The results from the aquaplanet simulations are not described here as they did not reach thermal equilibrium within the model timeframe.

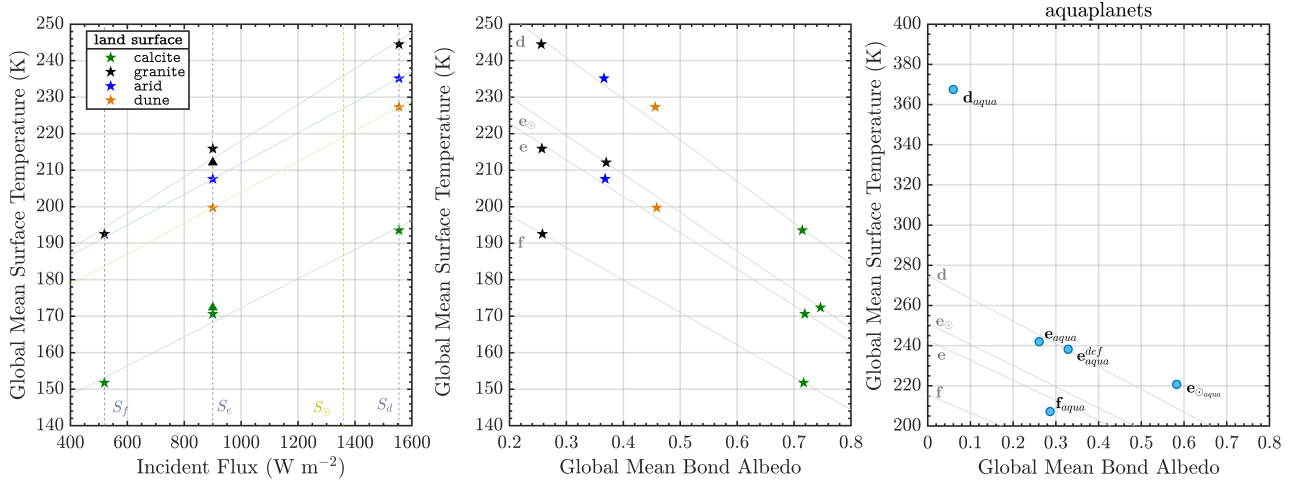

**Figure 2.** **Left**: Global mean surface temperature as a function of incident stellar flux for the different land surface types considered in this work (calcite: green marker, granite: black marker, aridisol: blue marker, dunesand: brown/red marker). The two simulations carried out for $e$ using a solar spectrum for input are denoted with a green and black triangle for calcite and granite scenarios, respectively. The flux levels for $d$, $e$, $f$, and Earth in the Sol system ($e_\odot$) are denoted with dashed lines. Solid gray lines denote linear best fits of incident flux/temperature (excluding the $e_\odot$ case). **Center**: Global mean surface temperature as a function of mean global albedo for the different land surface types considered in this work (calcite: green marker, granite: black marker, aridisol: blue marker, dunesand: brown/red marker). Solid gray lines denote linear best fits of albedo/temperature for the 4 land surface types for each planet. **Right**: Global mean surface temperature as a function of mean global albedo for the aquaplanet control cases, including planet $d$ and $e$ as well as $e$ using the CESM defaults for ice and snow ($e^{def}_{aqua}$), $e$ simulated with a solar spectrum ($e_{\odot aqua}$) and $f$. Solid gray lines denote linear best fits from the landplanet simulations (central panel) for comparison.

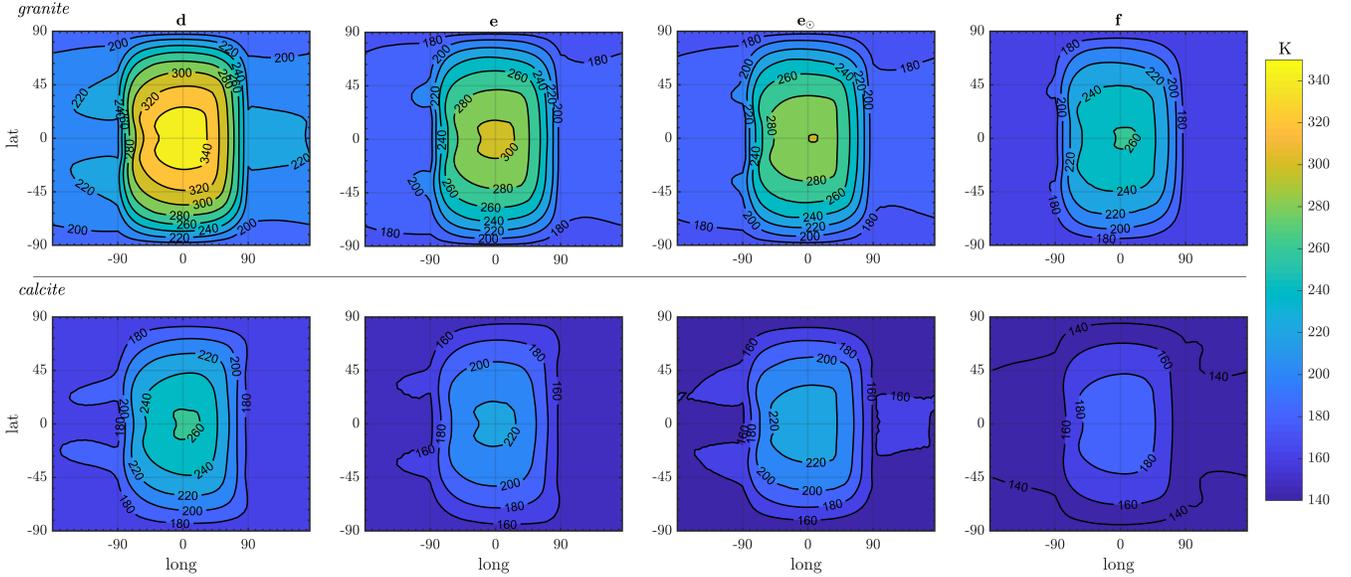

**Figure 3.** **Top row**: Contours (interval 20 K) of average surface temperature as a function of latitude and longitude for planet $d$, $e$, $e$ modelled using a solar spectrum ($e_\odot$), and $f$ for a homogeneous granite surface. **Bottow row**: Contours (interval 20 K) of surface temperature as a function of latitude and longitude for planet $d$, $e$, $e_\odot$, and $f$ for a homogeneous calcite surface.

Results for the globally averaged surface temperatures and albedos from the land-planet simulations are summarized in table 2, and presented in figures 2 and 3, as well as dayside (substellar hemisphere) to nightside (antistellar hemisphere), and equator to pole, temperature contrasts in figure 5. We emphasize when considering these results that these simulations were carried out in a dry atmosphere, and the addition of a water vapor source could significantly change the picture being described in this paper. We note the characteristic 'eyeball' of high surface temperatures slightly offset to the east of the substellar point particularly evident in the case of *d*, and the associated extension of somewhat higher equatorial surface temperatures into the nightside hemisphere through atmospheric heat transport (figure 3). As expected for synchronously rotating planets we note a strong dayside-nightside contrast in surface temperatures. On average, surface temperatures are higher on *d*, due to its smaller orbital separation and correspondingly higher instellation. However, our results show an 80 to 90 K difference in maximum annually-averaged surface temperatures at the substellar point between the granite (which exhibited the lowest average albedo) and calcite endmember (which had the highest albedo) cases (figure 3) for TRAPPIST-1d, $e$, and $f$, and a difference in globally-averaged surface temperature between the calcite and granite endmember cases of 45 to 50 K (figure 2). Therefore, we will most often contrast and compare the calcite and granite surfaces as they represent the upper (highest albedo) and lower (lowest albedo) bounds of our albedo parameter space, respectively, with the dune sand and aridisol results falling between these values.

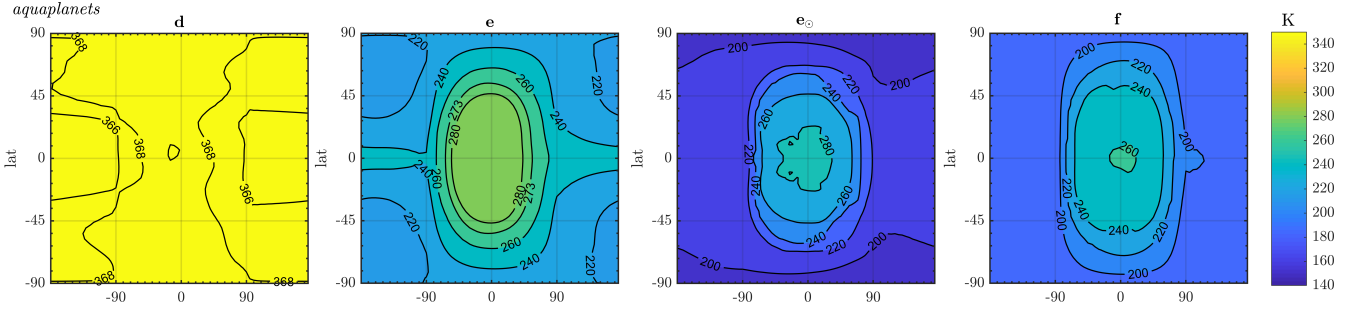

**Figure 4.** Contours (interval 20 K, except *d* where intervals are 2 K) of surface temperature as a function of latitude and longitude for aquaplanet *d*, *e*, $e_\odot$, and *f*.

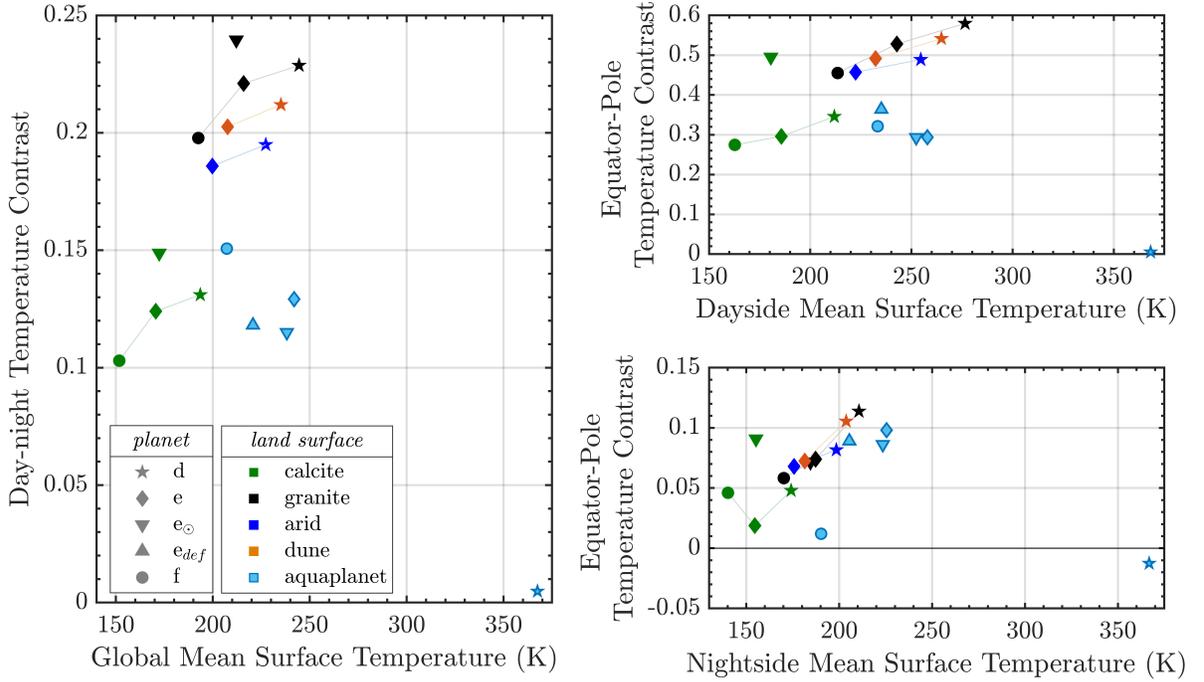

**Figure 5. Left**: Dayside-to-nightside temperature contrast ($[T_{day} - T_{night}]/T_{equil}$) (K) and global mean surface temperature (K). Icons denote planet *d*, *e*, *f*, planet *e* modelled using a solar spectrum ($e_\odot$), and planet *e* modelled using CESM's default snow and ice albedos ($e_{def}$) (only for aquaplanet simulations). Colors represent land surface type; calcite: green, granite: black, aridisol: dark blue, dunesand: brown/red, aquaplanet: light blue. **Top right**: Dayside equator-to-pole temperature contrast ($[T_{equator} - T_{polar}]/T_{equil}$) as a function of dayside mean surface temperature. **Bottom right**: Nightside equator-to-pole temperature contrast ($[T_{equator} - T_{polar}]/T_{equil}$) as a function of nightside mean surface temperature. Note the differing ranges of the y-axis between the figures on the right: the equator-to-pole temperature contrast is larger on the dayside than the nightside.

The high albedo of the calcite surface (0.714 for TRAPPIST-1d to 0.746 for TRAPPIST-1$e_\odot$) results in the lowest temperatures of four land surfaces explored here, the peak of which is 263 K (underneath the substellar point) for $d$, 226 K for $e$, and 194 K for $f$, with average dayside temperatures of 211, 185, and 162 K respectively (table 2). Conversely, the uniformly low albedo felsic surface of granite (∼0.25 for the non-Solar simulations) returns the highest surface temperatures. This can be compared with the current broadband albedo for the Earth given by Stephens (2015) as 0.29, Venus (0.75), and Mercury (0.119) (de Pater & Lissauer, 2001). For $d$ with a granitic land surface the surface temperatures peak at 353 K at the substellar point, while on planet $e$ and $f$ this peak is at 306 K and 261 K respectively. Average dayside temperatures range from 276 K ($d$) to 213 K ($f$). Despite receiving 14% more incident flux than the contemporary Earth (and assuming Earth-like levels of $CO_2$ and $CH_4$ greenhouse gases), and while exhibiting substellar temperatures on the order of 350 K, our results suggest that the *average* surface temperatures on the substellar/dayside hemispheres of TRAPPIST-1d are below freezing except the case of a granite land surface (when it is 3 K above). The high substellar temperatures yet relatively low average surface temperatures are due to the strong equator-pole temperature contrast associated with the granite landplanets (see figure 5), as well as variations in atmospheric heat transport induced by these gradients which we will discuss in a later section.

Both the dayside-to-nightside and equator-to-pole temperature contrast on synchronously-rotating planets such as those in the TRAPPIST-1 system can, to first order, elucidate heat and energy transport dynamics and provide a simple metric/diagnostic to investigate the effect of land surface composition and albedo on the transport of heat around the planet. Figure 5 displays both the day-night temperature contrast, that is the difference between the average surface temperature on the day side and that on the night side normalized to the equilibrium temperature ($[T_{day} - T_{night}]/T_{equil}$), as well as the equator-pole temperature contrast ($[T_{equator} - T_{polar}]/T_{equil}$) for both the dayside hemisphere and the nightside hemisphere. These data reveal a land surface composition/albedo dependence in that both the day-night and equator-pole (day and night side) temperature contrasts increase as the albedo of the land surface decreases. Therefore the granitic surface returns the largest day-night and equator-pole temperature contrasts for all four planets, and also that planets with higher surface temperatures have generally larger temperature contrasts. The spread in nightside temperatures, while large, is still substantially smaller than the spread in dayside temperatures.

Differences in climate and albedo between the planets modelled with the TRAPPIST-1 spectrum and that of the solar spectrum are also notable. We initialised calcite and granite landplanet simulations for planet $e$ using a Solar spectrum (henceforth $e_\odot$) but otherwise identical initial conditions. The solar spectrum, as it originated from a G-dwarf main-sequence star has a considerably different SED to that of TRAPPIST-1 (see figure 1) with a greater proportion of flux in shorter wavelengths. As a result, the calcite landplanet $e_\odot$ case was slightly warmer (+2 K), on average, than when using the TRAPPIST-1 spectrum, but the granite case is somewhat cooler (-3 K) (see figure 5, left), while the mean surface albedo increased in both cases from 0.72 to 0.75 and 0.26 to 0.37, respectively (see figure 2, center). The spectral effect on the climate and heat transport induced by the variation between the solar and TRAPPIST-1 spectrum is also evident in terms of the temperature contrast on the planets (figure 5) in which the $e_\odot$ case exhibits markedly greater day-night and equator-pole contrast than the TRAPPIST-1e case, independent of land surface type. For example, a granite-covered $e_\odot$ case returns the greatest day-night temperature contrast (0.23) of the planets considered here, as well as a dayside equator-pole contrast (0.57) equivalent to the much warmer planet $d$. This trend suggests that, for the same land surface composition and equivalent incident flux, planets orbiting lower mass stars will be slightly cooler while also exhibiting weaker gradients in surface temperature from the equatorial to polar regions, as well as from the day to the night hemispheres. This may indicate improved heat transport efficiency under the solar spectrum through zonal or meridional transport which we address in a later section.

The globally averaged surface temperatures of the aquaplanet simulations of planet $d$ and $e$ as well as $e$ using the CESM defaults for ice and snow ($e_{def}$), $e$ simulated with a solar spectrum ($e_\odot$), and planet $f$ (figures 2, right, 4, and

5) are generally higher than the landplanet configurations, and their day-night and equator-pole temperature contrasts similar to those of the calcite simulation. Higher surface temperatures are expected given the higher water vapor content of their atmospheres; water vapor is the dominant greenhouse gas on Earth today, and on planets in M dwarf systems it also is an important absorber of near-infrared radiation emitted by the star. The aquaplanet solutions for the TRAPPIST-1 planets considered here are stable, with the exception of TRAPPIST-1d which is in the process of entering a runaway greenhouse state here (figure 4) and did not reach equilibrium conditions when the model was stopped. Surface temperatures and albedo for the aquaplanet configuration of TRAPPIST-1d correspond to only the final month's data. At this point the planet has mean surface temperatures of ∼370 K, with a strong global inversion layer in troposphere approaching ∼450 K (e.g. Kopparapu et al. 2017, Fig 7), and persistent energy imbalance of ∼30 W m$^{-2}$. Note also that the day-night and equator-pole temperature contrasts are negligible (figure 5); in fact, the pole is slightly warmer than the equator on the nightside of the planet. We would expect this planet to continue to warm until it reached a terminal runaway greenhouse state, with temperatures approaching ∼1500 K (Goldblatt et al. 2013). This occurs when planetary heating results in the evaporation of surface water and oceanic reservoirs into the atmosphere initiating a positive feedback on specific humidity, surface temperature, and evaporation rate. The extremely low values of water in the landplanet simulations prohibit the occurrence of a runaway water vapour greenhouse in those simulations, though the presence of an ocean would modify that result.

### 3.2. Rotational regimes

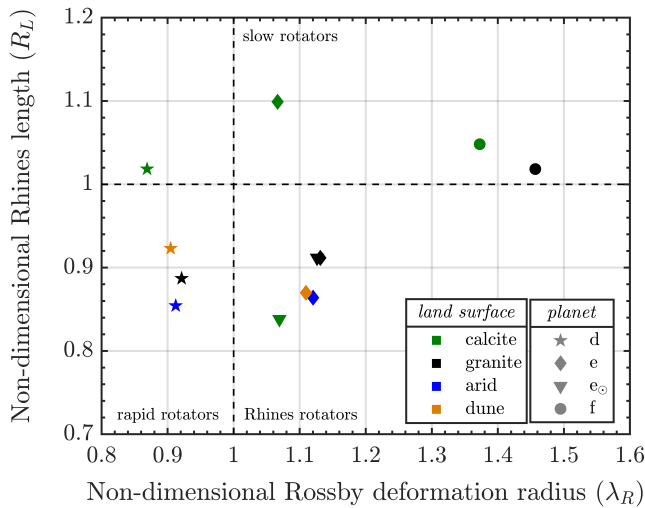

**Figure 6.** Rotational regimes for TRAPPIST-1 planets and land surfaces considered in this work. Tidally-locked planets are in the rapidly rotating regime when the non-dimensional Rossby deformation radius is less than one, which includes most land composition scenarios for TRAPPIST-1d (pentagram markers); TRAPPIST-1f (circle markers) falls into a slow rotator regime, and most simulations for TRAPPIST-1e suggest that the planet is a Rhines rotator, except in the case of the calcite simulation.

Synchronously rotating terrestrial planets, such as those in the tightly-packed TRAPPIST-1 system, exhibit markedly different planetary-scale atmospheric dynamics from asynchronous rotators, and can further be classified as slow rotators, rapid rotators, and Rhines rotators, based on the ratio between the Rossby deformation radius ($\lambda_R$) and Rhines length ($R_L$). The ratio between these parameters determines the circulatory regime of the atmosphere as a function of size of the planet and the scale of certain features in its atmosphere. The former parameter determines the maximum scale of vortex formation i.e. the scale at which circulatory features in the atmosphere are significantly affected by the planet's rotation and is defined as:

$$\lambda_R = \frac{\sqrt{gH}}{2\beta} \tag{1}$$

where $g$ is gravitational acceleration near the planet's surface, $H$ is the atmospheric scale height, and $\beta = 2\Omega/R_p$ (where $\Omega$ is the rotation rate ($2\pi$/rotation period) and $R_p$ is the planet radius) describes the equatorial Coriolis parameter (Showman *et al.* 2013; Haqq-Misra *et al.* 2018). The Rhines length ($R_L$) defines the scale at which planetary rotation induces zonal elongation of turbulent structures and is computed as $R_L = \pi\sqrt{U/\beta}$, where $U$ is the the area-weighted root mean squared surface wind on the day hemisphere (that is, the sum of the mean zonal and meridional components). **$R_L$ and $\lambda_R$ are then computed as non-dimensional ratios as a function of planetary radius ($R_p$) by dividing by $R_p$ and $R_p^2$, respectively** Furthermore, the Rossby deformation radius will be affected by average surface temperature ($T_S$) due to its affect on the atmospheric scale height ($H = T_s R_{gas}/m_{air}g$, where $R_{gas}$ is the gas constant (8.314 J K$^{-1}$ mol$^{-1}$) and $m_{air}$ the mean molecular mass of dry air (0.028 kg mol$^{-1}$)), and therefore the value of this parameter does change for each land surface type with implications for energy transport, climate stability, and detection.

In the case of Earth-sized terrestrial planets orbiting a cool star at low separations (on the order of ∼5-10 days), such as the TRAPPIST-1 planets considered in this work, Haqq-Misra *et al.* (2018) hypothesized that TRAPPIST-1d is likely in a rapid rotation regime, i.e. that the Rossby deformation radius is less than the planetary radius ($\lambda_R < R_P$), but that planets with wider orbital seperations likely fall into either the Rhines rotator or slow rotator regime. Figure 6 displays the **non-dimensional** $R_L/\lambda_R$ ratio for the TRAPPIST-1 system planets and land surface compositions considered here; our results do suggest that $d$ is a rapid rotator, $f$ is a slow rotator, and that $e$ is likely a Rhines rotator (except in the calcite simulation). Land surface composition does have a considerable effect on this ratio, as evident in the case of a calcite-covered TRAPPIST-1e, which falls into a slowly rotating regime. However, we note that the accurate determination of the scale height of the atmosphere and surface winds would allow for the rotational regime of the TRAPPIST-1 system planets to be better constrained.

### 3.3. *Heat and Energy Transport*

Following Lintner et al. (2004) and Laguë & Swann (2016) we compute zonally-averaged, cross-equatorial energy transport as a function of latitude ($\phi$) using the net radiation (the sum of the short- and longwave fluxes) at the top of the atmosphere (TOA), area-weighted to the radius of the planet. The poleward energy transport is the sum of energy transport at each latitude integrated from the south to north. Due to the very low water content of the land surface in our landplanet simulations, and the shallow 'slab' ocean configuration adopted in the aquaplanet cases, all energy transport in these simulations is attributable to the atmosphere. We use the 'energy flux potential' as a diagnostic to explore the energy balance and divergent heat transport around these planets, as shown in figure 7 (Trenberth et al. 2002; Boos & Korty, 2016). For each stellar spectrum considered, the granite land surface results in the most total shortwave energy absorbed by the system, while the calcite yields the least. Due to their synchronous rotation, it holds that energy is absorbed only on the day-side of the planet; while some of the energy is radiated away locally through outgoing longwave radiation (OLR) on the dayside of the planet, atmospheric circulation acts to move a large portion of the energy absorbed on the day-side of the planet to the nightside, where it can be efficiently radiated out of the system.

This difference in total energy input to the planet, either due to greater orbital seperation or different land surface compositions, results in markedly different strengths of atmospheric energy transport. To explore the 2-dimensional pattern of energy transport on these planets, we utilize the energy flux potential ($\chi$) (Trenberth et al. 2002; Boos & Korty, 2016). The energy flux potential quantifies the divergent or irrotational component of the zonal mean meridional atmospheric energy transport; energy flows from regions of low to high $\chi$ (vectors in figure 7) and the Laplacian of $\chi$ is equal to the seasonal mean net energy input to the atmospheric column (Boos & Korty, 2016). Energy is transported both zonally and meridionally by the atmosphere, away from the day-side of the planet towards the night-side of the planet. The energy flux potential and is greatest in the case of the granite land composition for planet $d$ (-7.4 PW at minimum $\chi$) suggesting vigorous heat transport from the substellar region both zonally along the equator and meridionally over the poles converging at or near the antistellar point on the nightside hemisphere. As granite yields the largest total energy absorbed by the system, the granite planets show the strongest divergent energy transport by the atmosphere, while the calcite planets show the weakest atmospheric energy transport.

#### 3.3.1. *Hadley Circulation*

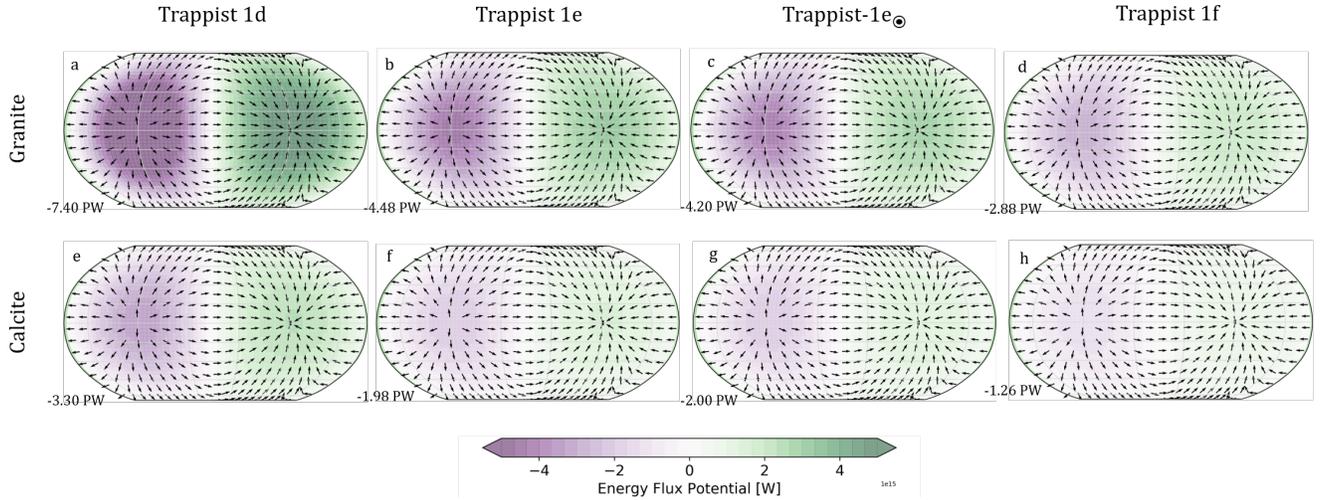

**Figure 7.** Energy Flux Potential ($\chi$) (W) for planets TRAPPIST-1d, e, e modelled with a solar spectrum, and TRAPPIST-1f using granite (top row) and calcite (bottom row) land surface compositions. Vectors are normalized divergent energy flux potential. Negative (purple) energy flux potential values occur on the day side of the planet, while positive (green) energy flux potential values occur on the night side of the planet. The minimum $\chi$ for each simulation is noted in the lower left of each panel (and in table 2)

At the land surface, the energy budget is closed over sufficiently long timescales, as the land has a small heat capacity and the model does not allow for lateral heat transport within the land surface. That is, all the shortwave (SW) and longwave (LW) energy absorbed by the land surface is emitted to the atmosphere either as LW radiation (as a function of surface temperature) or sensible heat. Latent heat is negligible given the dry nature of these systems. Thus, we can infer horizontal gradients of energy over the integrated atmospheric column by considering the pattern of the TOA energy imbalance. Gridcells with net energy into (out of) the atmospheric column at the TOA must necessarily have horizonal divergence (convergence) of energy within the atmospheric column. The TOA energy imbalance drives atmospheric energy transport both meridionally and zonally. We further explore the atmospheric circulations in the E-W and N-S directions.

In the aquaplanet simulations, each hemisphere has a single large overturning "Hadley" circulation (not shown, see Haqq-Misra et al. 2018; Showman et al. 2013). On the day-side, the circulation moves energy from the equator to the pole, while on the night side, the circulation moves energy from the pole towards the equator. However, on the land planets, the circulation is more complex (figure 8). We consider the meridional stream function which shows the zonally averaged mass transport, globally, for the day side only, and for the night-side only. Rather than a single overturning cell, each hemisphere generates both an equatorial and a polar overturning circulation. Averaged over all longitudes, the streamfunction collapses to a single cell moving energy away from the equator towards the poles. When considered separately, we see an equatorial cell on the day side that rises in the tropics and sinks in the subtropics. The polar cell sinks in the subtropics on the day side, but spans the pole, rising in the subtropics on the night side of the planet. The equatorial cell on the night side of the planet rises in the subtropics and sinks at the equator, opposite of the circulation on the day side. This opposing circulation necessarily requires a disconnect between the meridional overturning circulation on the day and night side of the planet.

### 3.3.2. *Walker Circulation*

We note a weak Walker-like circulation at the equator. Air rises under the substellar point, generally flows from west to east, and sinks on the night side of the planet. As with the meridional overturning circulation, the Walker-like circulation is weakest for the calcite land surface (figure 9). However, the strength of the zonal overturning circulation is less varied than the strength of the meridional overturning circulation between the other land surface types.

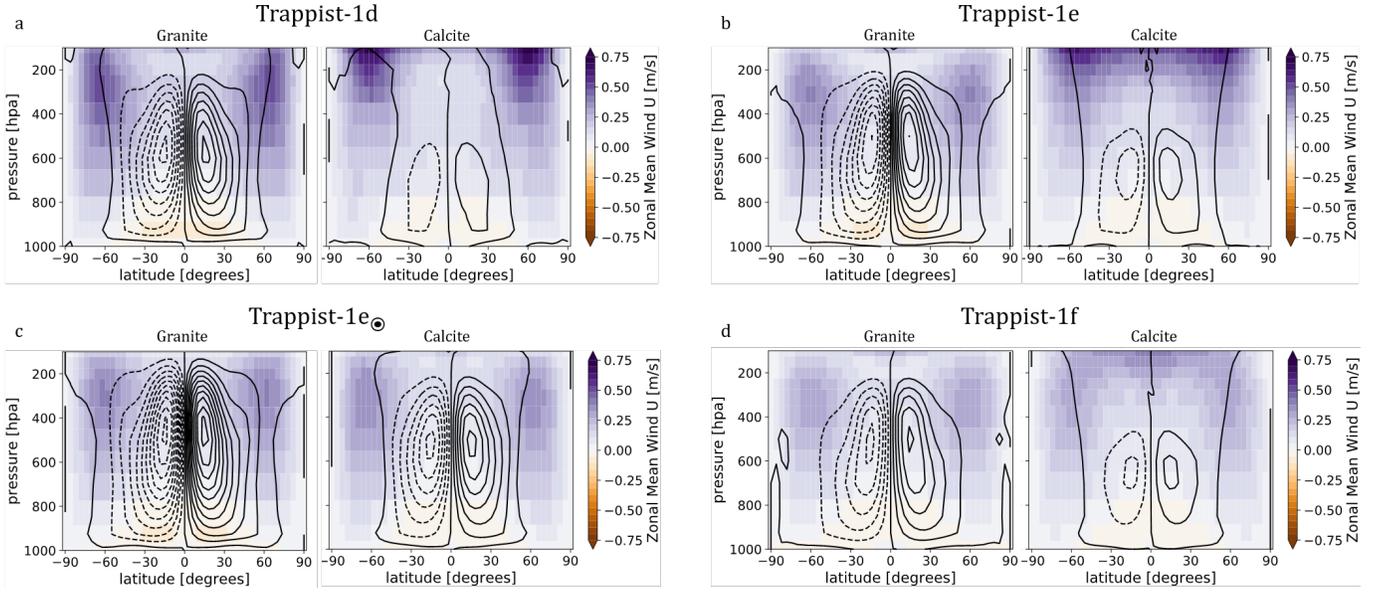

**Figure 8.** Mean meridional overturning circulation (kg s$^{-1}$) (Hadley cell) for homogeneous surface composition TRAPPIST-1 planets – granite (left), calcite (right). Contour intervals are 2.5 x 10$^{10}$ kg s$^{-1}$, with solid contours indicating clockwise and dashed contours indicating counter-clockwise circulation. Shading denotes zonal mean wind speed (m s$^{-1}$).

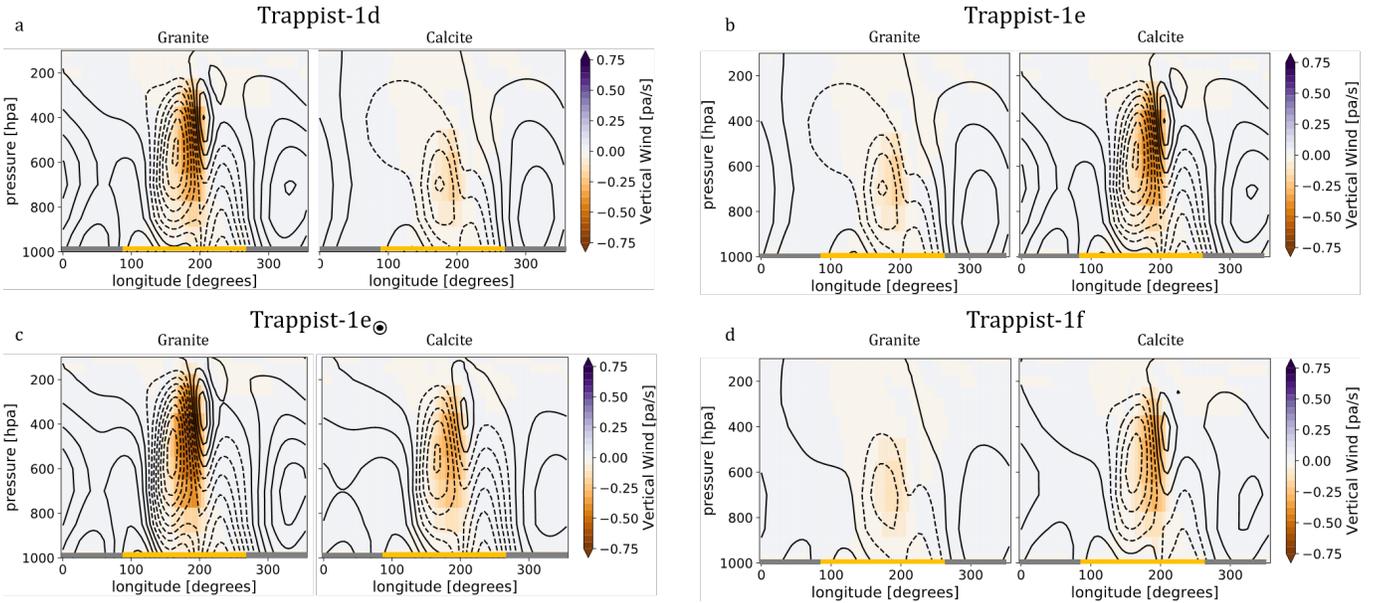

**Figure 9.** Global east-west overturning circulation at the equator (kg/s) (Walker circulation) for homogeneous surface composition TRAPPIST-1 planets – granite (left), calcite (right).Yellow shading denotes the dayside and gray the nightside (i.e., substellar centered at 180º) Contour intervals are 1 x 10$^{10}$ kg/s, with solid contours indicating clockwise and dashed contours indicating counter-clockwise circulation. Shading denotes vertical wind speed (pa/s), where positive values denote vertically upwards flow.

There are considerable differences in the strength of the zonal and meridional transport fluxes between the contrasting land surfaces. Here we can identify that, while the dayside hemisphere is responsible for more heat transport overall and that these fluxes are primarily zonal, the nightside nevertheless exhibits some polewards heat transport. Variations in the magnitude of both the meridional and zonal heat transport is evident between TRAPPIST-1d and e; however, the contrast between the planets is less than the differences in transport induced by variations in the land surface composition. TRAPPIST-1e, being cooler, has lower rates of overall heat transport than TRAPPIST-1d. However, we note the most significant strengthening of heat transport between the calcite (which also has the lowest surface temperatures) and the granite cases, and that the zone of greatest transport likely moves higher up in the atmosphere with the greater surface temperatures associated with the uniform, low albedo granite surface. As the strength of the overturning circulation is not directly related to the strength of its mass transport (as gross moist stability is not necessarily constant across all climates), and by focusing on the energy flux potential, we can determine that energy is transported in that column but not precisely at which level. However, we hypothesize that, due to the very low water vapor content of the atmosphere, mass and energy transport are more closely related for landplanets than they may be under moist conditions.

Significantly, we identify evidence of a rapidly rotating circulation that spans less than a hemisphere in the calcite simulations of $d$, $e$, and $f$ (figure 9), a feature that is diagnostic of a rapid rotator as initially suggested by the $R_L/\lambda_R$ ratio presented in figure 6, but not in the case of TRAPPIST-1e when simulated with a solar spectrum ($e_\odot$) (figure 9c); here, we can identify a super-rotating cell in the upper atmosphere, which is also evident in the granite surface simulations for $e$, that is indicative of a Rhines rotator. Similar to the slowly rotating TRAPPIST-1f (figure 9d) Rhines rotators exhibit day to night-side Walker circulations, but also asymmetric zonal jets in the mid-latitudes. These features are identifiable in figure 8b, 8c as a sharp gradient in the strength of the northwards clockwise rotation in the upper atmosphere, but at lower latitudes which may be a result of the homogenous land surface increasing the equator-to-pole temperature contrast.

### 3.3.3. *Winds*

At the surface, winds flow towards the substellar point on the day side of the planet (figure 10). This flow of wind up the energy gradient is driven by the strong dry convection directly beneath the substellar point. Winds flow towards the zone of peak insolation from all directions (east, west, north, and south). Surface winds diverge from two points in the high latitudes on the night side of the planet, just west of the western terminator, flowing equatorward and day-ward from that point. Polewards of the antistellar point on the night side, surface air flows up over the pole to reach the dayside following a generally north-south line of flow, while equatorwards of the antistellar point, surface air flows equatorward and turns to follow a generally east-west line of flow. In most of our simulations, the planets have two high latitude jets which exist on both the day and night sides of the planet. The jets veer equatorwards in the eastern half of the dayside, then turn polewards again once they reach the night side. For planet $e$, winds aloft (250 hPa) flow eastward along the equator on both the day and night sides of the planet for the calcite simulation, as expected in the case of a slow rotator. Recall that the slowly rotating calcite TRAPPIST-1e, as well as the granite and calcite $f$ scenarios, have the weakest dry convection at the substellar point. In contrast, the other land types (granite, dunesand, and aridisol) all show divergent winds at the equator on the day side of the planet, resulting from strong rising motion associated with Rhines rotators. This displacement causes the day-side jets to veer equatorward after the substellar point. On the night side of the land planets, the winds are strongest in the jets which are located in the mid-to-high latitudes.

### 3.3.4. *Atmospheric Heating*

We are able to summarize the contribution of incident flux and heat transport to the overall heat budget of the atmosphere by considering the atmospheric 'radiational heating' ($Q_R$) component of our results, which we generate from the combination of the rate of heating due to atmospheric absorption of stellar flux ($Q_{RS}$) and the rate of heating or cooling due to longwave radiation ($Q_{RL}$) so that $Q_R = Q_{RS} + Q_{RL}$ (Cheng et al. 1977). We present in figure 11 these hemispherically-averaged (that is, over the substellar and antistellar hemispheres) data for the lower atmosphere (>900 hPa). These results support and further illustrate the transport mechanisms highlighted in the previous section; specifically, the zonal transport of heat in the mid-latitudes is evident, contributing to a net positive heating rate along 45 degrees N and S, along with a zone of eastward flow above the equator. While some complex thermal structure is evident in the atmosphere above the mid-latitudes and poles in most cases, the atmosphere above ∼940 hPa provides a net cooling (i.e. LW emission) in general. While the trend holds for most land surface compositions considered here, the strength of radiational heating declines, or rather the difference between the rate of incident flux and longwave emission from the atmosphere is reduced, as global average surface temperature declines and therefore the granite-covered TRAPPIST-1d exhibits the greatest rates of radiational heating and most vigorous heat transport. Differences in atmospheric heating as a function of rotational regime are also evident from figure 11, in which the slow rotators (TRAPPIST-1f, and TRAPPIST-1e calcite) exhibit weaker vertical heating gradients – suggesting day to night-side Walker circulation and zonal heat transport along these isotherms – than the Rhines and (TRAPPIST-1$e_\odot$) and rapid rotators (TRAPPIST-1d) in which strong zonal jets transport heat along the mid-latitudes.

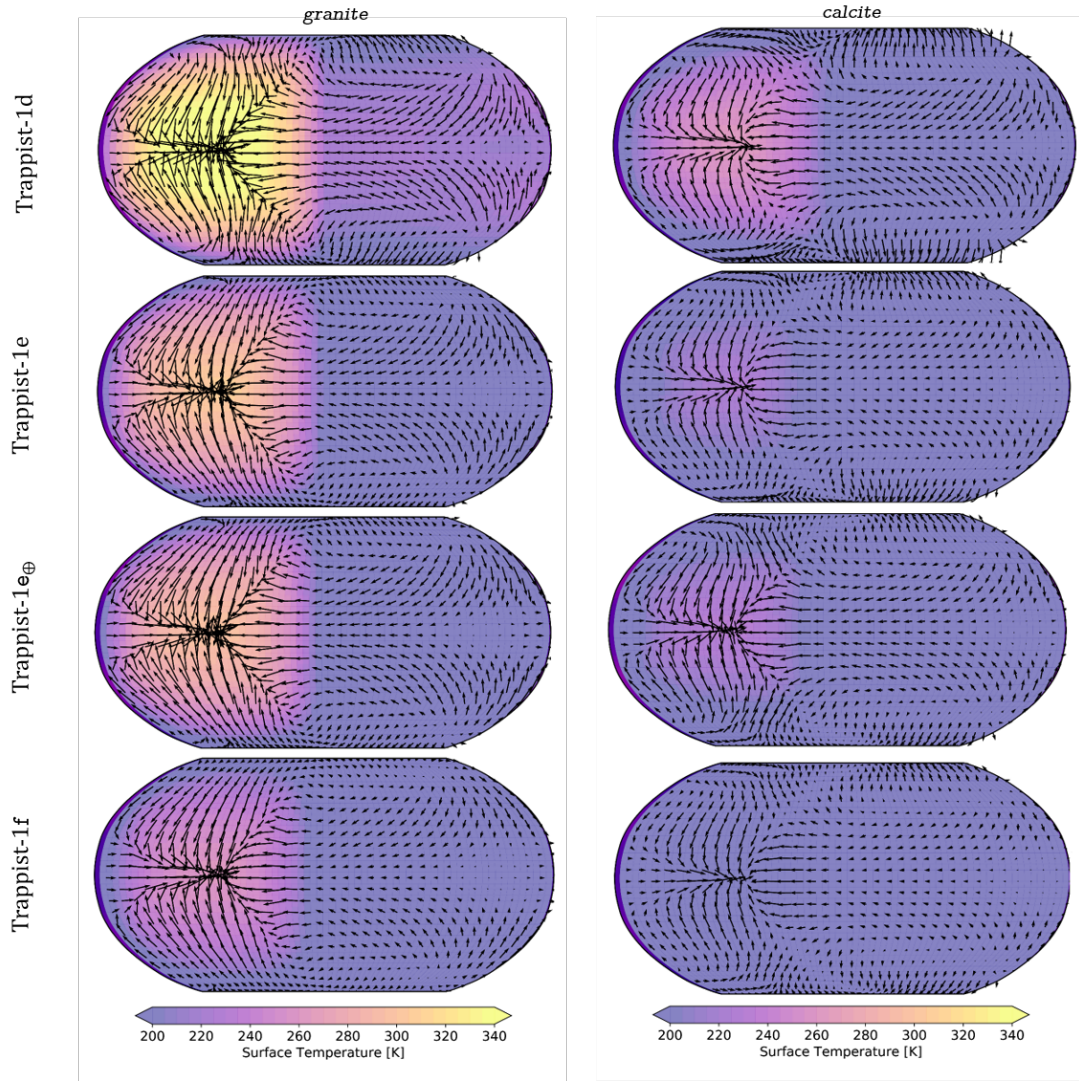

**Figure 10.** Surface winds (vectors of direction and magnitude) and contours of surface temperature (K) (color) for TRAPPIST-1d (top row), $e$, $e_\odot$, and $f$ (bottom row) for granite (left column) and calcite (right column) land surface compositions.

## 4. DISCUSSION

The results presented here suggest that changes in albedo due to varying land surface composition have a considerable effect on the climate and atmospheric circulation of synchronously-rotating, desiccated landplanets. In the case of the TRAPPIST-1 system in particular, the interior of the three planets considered here, TRAPPIST-1d, may be able to support zonal average surface temperatures that are above the freezing point of water on much of the dayside hemisphere, if the surface is comprised of a relatively low albedo material such as an igneous rock surface. As these simulations are carried out with the assumption of negligible water content in the land and atmospheric reservoirs, the temperature and heat transport fluxes we present here are likely conservative. The addition of small amounts of water into these otherwise land-covered worlds would likely provide further interesting climate feedbacks between a water-vapor greenhouse and cold trapping (e.g. Abe et al. 2011; Leconte et al. 2013). In these habitable "dune" like worlds, the albedo of the land surface may be critical for determining the fate of water present on these worlds.

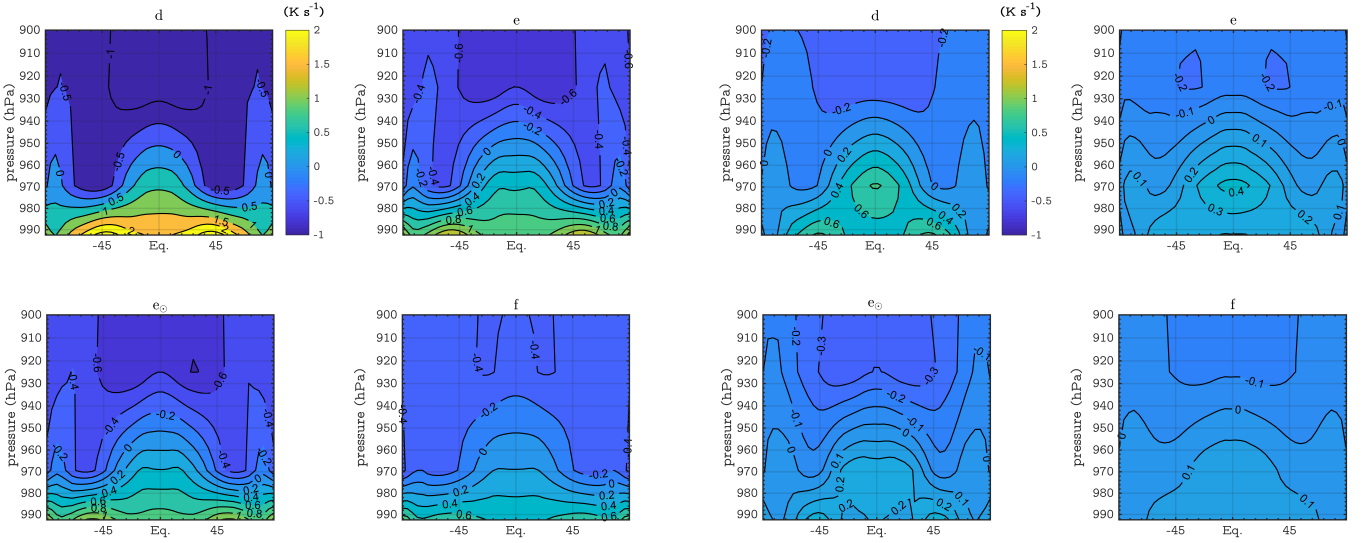

(a) Granite surface composition

(b) Calcite surface composition

**Figure 11.** Globally-averaged contours of atmospheric heating ($Q_R$) (K s$^{-1}$) as a function of latitude and pressure (lower atmosphere (>900 hPa) plotted) for granite surfaces (left two columns (a)) and calcite surfaces (right two columns (b)) on TRAPPIST-1d (top left), e (top right), $e_\odot$ (bottom left), and f (bottom right).

Alternatively, an ocean-covered aquaplanet configuration produces the characteristic symmetrical 'eyeball' pattern of an area of unfrozen, open ocean at the substellar point and surface temperatures considerably higher than in the land-only cases (figure 5). These results are in general agreement with other workers (e.g. Checlair et al. 2019), but we acknowledge that other factors such as the amount of CO$_2$ outgassing and, in the case of non-aquaplanet configurations, the distribution of any continental landmasses, will exert some control on the substellar unglaciated zone. Recall, also, that these data are from only the last month of the ∼90 year simulation as these aquaplanet simulations failed to reach thermal equilibrium. The differences in surface temperature and global mean albedo between the three TRAPPIST-1e aquaplanet scenarios is contingent on the differences in albedo of ice and snow used in our TRAPPIST-1 and the CESM default simulations, and, in the case of $e_{\odot aqua}$, between the spectral energy distribution of TRAPPIST-1 and Sol. The $e_{\odot aqua}$ case is cooler, on average, and has a higher albedo as water ice appears more reflective in longer wavelengths that are associated with **hotter stars like the Sun** (Rushby et al., 2019; Shields et al., 2013).

The large surface temperature variations, as well as day-night and equator-pole temperature contrasts, observed in the simulations of TRAPPIST-1d reveal the effect of the land albedo on planetary climate. With a uniformly high albedo, land-covered configuration returns a maximum zonally averaged mean surface temperature ∼50 K cooler than that of a lower albedo surface. A calcite-covered TRAPPIST-1d configuration exhibits comparable surface temperatures to TRAPPIST-1f with a granite land surface, despite receiving three times the incident stellar flux. It is unlikely that any land-covered planet will be spatially homogeneous in terms of surface composition and albedo, but we present end-member cases here using empirical spectra of terrestrial minerals in order to demonstrate the response of the climate system to changes in land albedo induced by compositional variation and continental configuration. In the case of a TRAPPIST-1d aquaplanet configuration the corresponding increases in the absorption of incident radiation and the higher heat capacity of the surface, coupled with the greater transport fluxes facilitated by the moist atmosphere, result in a runaway greenhouse feedback at its current level of instellation ($S_\oplus = 1.143$) (figure 2 (right)). Very high surface temperatures, as well as the dense opaque atmosphere, would preclude the potential to support an Earth-like biosphere. Therefore, we demonstrate that determining the composition and albedo of any potential land surface, as well as the distribution and fractional coverage of continental landmasses, will be crucial in making accurate determinations of the climate and potential habitability of terrestrial exoplanets.

Our results regarding zonal and meridional heat transport on synchronously-rotating landplanets suggest that the day hemisphere is moving considerable energy both poleward and meridionally over to the night side, with a much larger gradient east-west than north-south, as would be expected in a spin-orbit locked configuration. However, the antistellar hemisphere remains dynamic, and zonal and meridional heat transport does occur on the nightside, albeit at a lower rate than the substellar hemisphere. We also note a region of convergence in the mid-latitudes of the unlit hemisphere, which contrasts with both the asynchronously-rotating contemporary Earth, as well as the spin-orbit locked aquaplanet configurations, in that the primary zone of convergence in these cases is at the poles with zonal and meridional transport conforming to Walker-like circulation. We see much greater and more vigorous zonal and meridional transport fluxes on planets that have higher average surface temperatures, and therefore the lowest albedo land compositions (such as granite) induce the greatest amounts of atmospheric heat transport. This may have implications for the rotation of these planets as Leconte et al. (2015) suggest that even a relatively diffuse atmosphere may generate sufficient thermal tides to induce asynchronous rotation. Furthermore, Leconte (2018) argues that gravitational tides exerted on tidally-locked planets may reorient the mantle and crust such that continents would be aligned at the either the substellar and/or antistellar points, thereby emphasizing the need to understand the effects on albedo and climate of variations in land surface composition.

We note that the mean zonal circulation generated on the rapid rotator $d$ does not reach from the day to the night side and is dissipated by atmospheric dynamics on a scale much less than that of the planetary radius, as well as relatively weak convection (for the level of incident flux) beneath the substellar point from our energy flux potential results (figure 7, figure 9). While land surface composition does have an effect on the Rhines length (due to its effect on surface temperature from which the atmospheric scale height is derived), three of the four compositional scenarios considered in this work for TRAPPIST-1d put the planet in the rapidly rotating regime. A calcite land surface on $d$ suggests $R_L/Rp \geq 1$ and circulatory features more closely associated with slowly rotating planets are evident (figure 9 (right)), but we note that future atmospheric characterization of this planet could reduce the uncertainties in determining the surface wind speeds and atmospheric scale height and better constrain the $R_L/\lambda_R$ ratio. If both the Rhines length and Rossby deformation radius are greater than unity, the planet would be in a slow rotator regime, as in the case of TRAPPIST-1f in both the granite and calcite scenarios. Slow rotators are more likely to exhibit strong day-night transport through Rossby and Kelvin waves, and convective motion beneath the substellar point, but the low temperatures and incident flux associated with a planet at the orbital separation of $f$ make these atmospheric dynamics energy-limited and comparisons between the warmer planets in the system difficult.

Except in the calcite case, TRAPPIST-1e appears to be a Rhines rotator as its non-dimensional Rossby deformation radius is greater than the planet radius ($\lambda_R/R_p > 1$), but its Rhines length is less than one. Rhines rotators represent a transitional configuration between slow and rapid synchronous rotators, and are characterized by powerful super-rotating features in the upper atmosphere (figure 9(left)) coupled with significant upwelling at the substellar point (figure 7 ($e$, top row)) (Haqq-Misra et al. 2018). Consistent with our calcite simulations of TRAPPIST-1e being a slow rotator, or a Rhines rotator in the case of the other land surface compositions, (figure 6) we see the largest gradients in energy flux potential ($\chi$) in our $e$ and $e_\odot$ simulations as vigorous convective motion beneath the substellar point generates strong, thermally direct circulation with air heating and rising (negative $\chi$) on the dayside hemisphere and sinking and cooling (postive $\chi$) on the nightside (figure 7). However, we note an albedo/land surface composition effect on these results, as our calcite simulation of planet $e$ suggests a slow rotator, and furthermore also a spectral effect, in that our simulation of TRAPPIST-1e with a solar spectrum ($e_\odot$) returns a lower $R_L$ and puts the planet back into the Rhines regime. Future observations, or attempts at atmospheric spectroscopy and characterization, would reduce uncertainty in terms of wind speeds and atmospheric scale height, but these model results nevertheless illustrate that $e$'s atmospheric circulatory regime is contingent on the composition of the land surface and its distribution on the surface of the planet, and that variations in these parameters can have a considerable effect on heat transport around the planet. The variation in energy flux potential as a function of land surface type for this planet suggests a non-linear response between atmospheric heat transport and surface composition, in which lower albedo surfaces generate greater equator-pole and day-night surface temperature contrasts, which in turn induce stronger substellar convective motion and the formation of super-rotating features in the upper atmosphere that more efficiently distribute heat around the planet. This is significant as the TRAPPIST-1 system represents a robust natural laboratory of synchronously-rotating terrestrial planets that may fall into the rapid, Rhines, and/or slowly rotating regime primarily depending on orbital separation, but also contingent on the composition and distribution of land on their surfaces, allowing for further comparative planetology within the system, and between other terrestrial planet-hosting stars.

We have sought to carry out an investigation into the effect of land albedo on the climate of three Earth-sized terrestrial planets in the TRAPPIST-1 system and in so doing have assumed that these planets are uniformly land covered, and that the land surface is topographically featureless and composed of a single mineral or assemblage, and that the atmosphere of these planets hold 'Earth-like' quantities of climatically important greenhouse gases such as $CO_2$ and $CH_4$ while also (unlike Earth) being desiccated. We anticipate that the addition of water vapor will have a significant effect on these results. For example, 3D GCM results from Kodama et al. (2019) reveal that the instellation that a planet receives increases with an increase of the dry surface area. Furthermore, Lewis et al. (2018) demonstrate that for small substellar continents the land area could be completely covered by clouds, and thus the surface albedo effect could be muted. However, that paper also shows that for significant size substellar super-continent, the interior continent may remain drier with fewer clouds, allowing the surface albedo to potentially play a bigger role (Lewis et al. 2018). More work needs to be done to examine the effect of the surface albedo when some water is also present. We have justified these assumptions elsewhere in the paper, but note here that we aimed to strike a balance between computational efficiency, given the considerable computing time and resources required of these coupled simulations, and a wide albedo/land surface composition parameter space coverage. While unrealistic to assume an entirely homogeneous composition of the continental crust, we also recognize the inherent difficulties in simulating a sufficiently broad suite of possible compositional assemblages comprised of mixed materials for terrestrial planets of which we have no direct observations or indications of the chemical properties of their crust, and therefore we opted to bound our investigation by selecting compositions that are both present on Earth and exhibit a sufficient variability in Bond albedo. Future work in this area would incorporate a variable land/ocean ratio and dynamic hydrological cycle, building on the results from Rushby et al. (2019), Leconte et al. (2013), Leconte (2015, and Lewis et al. (2018), as well as topographic variability and mixtures of possible land surface compositions, to better constrain shifts of patterns of atmospheric circulation, rotational structures, and long-term, tidally-induced planetary reorientation.

## 5. CONCLUSION

We have investigated the effect of the composition of the land surface on the climate of land-dominated planets orbiting small, cool stars, using three Earth-sized terrestrial planets in the TRAPPIST-1 system as case studies. The climate of TRAPPIST-1d, *e*, and *f*, represented here by annually averaged surface temperatures and atmospheric heat transport fluxes, are strongly affected by changes in the albedo of the land surface. Assuming a homogeneous, completely land-covered configuration for each planet reveals that, should the land surface be primarily comprised of materials with higher albedo in the infrared (i.e. calcite), we observe a difference of ∼50 K in surface temperatures and a reduction in rates of cross-equatorial heat transport relative to a more absorptive crustal material such as granite or dry, weathered soils. Furthermore, we demonstrate that land surface composition has a control over the potential rotational regime of the atmosphere of terrestrial planets, and TRAPPIST-1e in particular, with lower albedo surfaces inducing stronger temperature contrasts between the equatorial and polar regions and the substellar and antistellar hemispheres that results in a **greater transport** from the day to the nightside of the planet, increased substellar convective uplift, and upper atmosphere super-rotation. When comparing these results to that of aquaplanet controls, we also note a considerable difference in climate response. An aquaplanet configuration for TRAPPIST-1d results in an unstable, runaway greenhouse regime with very high surface temperatures. Conversely, a calcite-dominated landplanet at the same orbital separation/instellation returns an annual average surface temperature approximately 160 K less than the aquaplanet case (228 K versus 367 K), demonstrating the importance of knowledge of the extent, distribution, and composition of the land surface of any terrestrial planet when considering its climate. However, we note that this work makes a small stride towards identifying the complex climatic feedbacks that may occur due to land surface changes on terrestrial planets, but additional work is needed to further constrain, in particular, the combined effects of atmospheric water vapour, land surface composition and distribution, and long-term geological and tidally-induced orbital evolution.

ACKNOWLEDGMENTS

This material is based upon work supported by the National Science Foundation under Award No. 1753373, and by NASA under grant number NNH16ZDA001N, which is part of the NASA "Habitable Worlds" program. The CESM project is supported primarily by the National Science Foundation. Computing resources were provided by the Climate Simulation Laboratory at NCAR's Computational and Information Systems Laboratory (CISL). MML acknowledges support from the James S. McDonnell Foundation. E.T.W. acknowledges support by the NASA Astrobiology Program Grant Number 80NSSC18K0829 and benefited from participation in the NASA Nexus for Exoplanet Systems Science (NExSS) research coordination network.

*Facilities:* Discover Supercomputer (NASA Center for Climate Simulation)

*Software:* CESM (NSF), ExoCAM

REFERENCES

[1] Abe, Y. *et al.* (2011), *Astrobio* 11 (**5**)
[2] Baldridge, A. M. *et al.* (2009), *Remote Sens. Environ.* 13
[3] Boos, W. R. & Korty, R. L. (2016), *Nature Geo* 9
[4] Chance, K. & Kurucz (2010), *J Quant Spectrosc Radiat Transf* (**111**)
[5] Checlair, J. H. *et al.* (2019), *ApJL* 884 (**2**)
[6] Chang, J. (ed) (1977), General Circulation Models of the Atmosphere (New York, NY: Academic Press, Inc.)
[7] Cockell, C. S. *et al.* (2016), *Astrobio* 16 (**1**)
[8] de Pater, I. & and Lissauer, J. *Planetary Sciences*, Cambridge University Press, 2001.
[9] Edson, A. R. *et al.* (2012), *Astrobio* 12 (**6**)
[10] Foley, B. J. (2015), *ApJ* 812 (**36**)
[11] Gillon, M. *et al.* (2017), *Nature* 542 (**456**)
[12] Goldblatt, C. *et al.* (2013), *Nature Geo* 6 (661-667)
[13] Greene, T. P. *et al.* (2016), *ApJ* 817 (**17**)
[14] Grimm, S. L. *et al.* (2018), *A&A* 613 (**A68**)
[15] Haqq-Misra, J. *et al.* (2018), *ApJ* 852 (**67**)
[16] Kalirai, J. (2018), *ConPh* 59 (**251**)
[17] Kiang, N. Y. *et al.* (2018), *Astrobio* 18 (**6**)
[18] Kodama, T. *et al.* (2019), *JGR Planets* 124 (**8**)
[19] Kopparapu, R. K. *et al.* (2017), *ApJ* 845 (**5**)
[20] Laguë, M. M. & Swann, A. L. S (2016), *JClim* 15 (**29**)
[21] Leconte, J. R. *et al.* (2013), *A&A* 554 (**A69**)
[22] Leconte, J. R. *et al.* (2015), *Science* 347 (**6222**)
[23] Leconte, J. R. (2018), *Nature Geo* 2018 (**11**)
[24] Lewis, N. T. *et al.* (2018), *ApJ* 854 (**2**)
[25] Lincowski, A. P. *et al.* (2018), *ApJ* 867 (**1**)
[26] Lintner, B. R. *et al.* (2004), *JGR Atmos* 109 (**13**)
[27] Meerdink, S. K. *et al.* (2019), *Remote Sens. Environ.* 230 (**111196**)
[28] Poulsen, C. J. *et al.* (2001), *Geophys. Res. Lett.* 28 (**8**)
[29] Pierrehumbert, R. T. (2011), *ApJL* 726 (**8**)
[30] Ricker, G.R. *et al.* (2014), *J. of Astron. Telescopes, Instruments, & Systems* 1 (**1**) 014003
[31] Rushby, A. J. *et al.* (2018), *Astrobio* 18 (**5**)
[32] Rushby, A. J. *et al.* (2019), *ApJ* 87 (**29**)
[33] Segura, A. *et al.* (2003), *Astrobio* 3 (**4**)
[34] Shields, A. L. *et al.* (2013), *Astrobio* 13 (**8**)
[35] Shields, A. L. *et al.* (2014), *ApJL* 785 (**1**)
[36] Shields, A. L. & Carns, R. C. (2018), *ApJ* 87 (**11**)
[37] Shields, A. L. (2019), *ApJS* 243 (**30**)
[38] Showman, A. P. *et al.* (2013) in Comparative Climatology of Terrestrial Planets, ed. S. J. Mackwell *et al.* (Tucson, AZ: Univ. of Arizona)
[39] Stephens, G.L. *et al.* (2015), *Rev.Geophys* **53**
[40] Trenberth, K. E. *et al.* (2002), *JGR Atmos* 107 (**D8**)
[41] Turbet, M. *et al.* (2018), *A&A* 612 (**A86**)
[42] Unterborn, C. T. *et al.* (2018), *RNAAS* 2 (**3**)
[43] Walker, J. C. G. *et al.* (1981), *JGR* 86 (**C10**)
[44] Wolf, E. T. (2017), *ApJL* 839 (**L1**)
[45] Wolf, E. T. (2018), *ApJL* 855 (**L14**)